\documentclass[reprint,prd,tightenlines,nofootinbib]{revtex4-2}
\usepackage{hyperref}
\hypersetup{colorlinks,allcolors=black}
 \usepackage{graphicx}
 \usepackage{float}
 \usepackage{amsmath}
 \usepackage{amssymb}
 \usepackage[utf8]{inputenc}
 \usepackage{natbib}
 \usepackage{subfigure}
 \usepackage[normalem]{ulem}
 \usepackage{cancel}
 \usepackage{soul}
 \usepackage{xcolor}
 \begin{document}
\title{The non-standard Lagrangian from non-uniqueness principle of the real scalar field and fermion field}
	\author{Suppanat Supanyo$^{1,2}$}
 	\author{Monsit Tanasittikosol$^{1,2}$}
 	\author{Sikarin Yoo-Kong$^3$}
 		\affiliation{$^1$Theoretical and Computational Physics (TCP) Group, Department of Physics, Faculty of Science, King Mongkut's University of Technology Thonburi, Bangkok 10140, Thailand}
 		\affiliation{$^2$Theoretical and Computational Science Centre (TaCS), Faculty of Science, King Mongkut's University of Technology Thonburi, Bangkok 10140, Thailand}
 	\affiliation{$^3$The institute for Fundamental Study (IF), Naresuan University, Phitsanulok 65000, Thailand}
 		\begin{abstract}
   We construct the non-standard Lagrangian, called the multiplicative form, of the homogeneous scalar field and fermion field through the inverse calculus of variations, which the equation of motion still satisfies the Klein-Gordon and Dirac equations, respectively. By employing the non-uniqueness of Lagrangian, we show that the Lagrangians can be written between the linear combination of standard and non-standard Lagrangian. The stability of the ghost field, an unnatural smallness of cosmological constant, and the chiral condensate are discussed by applying these new Lagrangians.
\end{abstract}
\maketitle
\section{Introduction}
The non-standard Lagrangian is another approach to explain unsolved problems in modern physics without concerning the renormalizability. In this point of view, the standard renormalizable Lagrangian, which consists of operators with mass dimensions less than four,  is assumed to be the lowest approximation from the ultraviolet (UV) complete theory.
As a consequence, the non-standard kinetic energy and potential energy are free to modify the UV physics. The construction schemes of the non-standard Lagrangian are categorized into two approaches: the top-down and bottom-up perspective. From the top-down perspective, for example, the Born-Infeld Lagrangian\cite{born1934foundations},  in which kinetic energy itself is introduced in square root as the relativistic Lagrangian, can be obtained in an appropriate limit of the string theory\cite{callan1998brane,gibbons1998born,Alishahiha:2004eh}. From the bottom-up perspective, the non-standard form of Lagrangian can be an arbitrary function while preserving the symmetry of the model and satisfying the experimental observation. Frequently, several forms of kinetic energy and potential energy are artificially introduced  to explain the new behaviors without systematically mathematical construction, such as the k-essence model \cite{PhysRevLett.85.4438, PhysRevD.63.103510, PhysRevD.103.043518}. According to this perspective, this framework is identical to the effective field theory from the bottom-up perspective \cite{BECHTLE2022129,Brivio:2017vri,PhysRevD.104.015026,PhysRevLett.121.111801} to generate the new appropriate interaction.  

 ``The non-uniqueness of Lagrangian''\cite{mizel1995nonuniqueness} and ``the inverse problem of the calculus of variations''\cite{douglas1941solution, hojman1981inverse, sarlet1982helmholtz} are another way to provide the non-standard form of Lagrangian. 
In the one degree of freedom system, the form of Lagrangian can be an arbitrary appropriate function instead of $T-V$ while the equation of motion (EOM) remains the same form, where $T=m\dot{x}^2/2$. For example, the quartic Lagrangian 
\begin{align}\label{LMbig1}
   L= \frac{T^2}{3}+2TV-V^2,
\end{align}
and the multiplicative Lagrangian \cite{surawuttinack2016multiplicative} 
\begin{align}\label{LMbig2}
   L= m\lambda^2\left(e^{-\frac{\dot{x}^2}{2\lambda^2}}+\frac{\dot{x}}{\lambda^2}\int_0^{\dot{x}} e^{-\frac{v^2}{2\lambda^2}}dv\right)e^{-\frac{V}{m\lambda^2}},
\end{align}
are alternative descriptions of Newton's law of motion
\begin{align}
    m\ddot{x}=-\frac{d V}{d x}.
\end{align}
Here, $\lambda$ is a parameter with the unit of velocity.
 These types of non-standard Lagrangians are not broadly discussed in \sout{the} classical mechanics, since this modified Lagrangian does not lead to any effect in the classical EOM. Hence, working with standard form is more pleasant. However,  the non-standard Lagrangians can give rise to the modified energy function, which leads to the different quantum theory \cite{henneaux1982lagrangians, PhysRevD.21.418}. One of the unsolved problems related to the non-uniqueness of Lagrangian is the strong-CP problem \cite{PhysRevD.86.036002, Pendlebury:2015lrz,Wurm:2019yfj}. The strong CP violation term with arbitrary phase constant is allowed in the Lagrangian without the effect on the classical EOM but this term can give rise to the unnatural small quantum effect observed through neutron electric dipole moment \cite{PhysRevD.86.036002, Pendlebury:2015lrz,Wurm:2019yfj}. In our hypothesis, the term in Lagrangian, which vanishes in classical EOM, might exist in the other types of fields such as scalar or fermion fields leading to a non-standard form of Lagrangian. However, the non-uniqueness of Lagrangian in the field theory has never been intensively mentioned because of various problems as follows.

$\bullet$ Renormalizablilty: If the Lagrangian of scalar or fermion fields is in non-standard form, the non-renormalizable term could exist. The renormalization is no longer applicable. Therefore, the non-perturbative method as the lattice calculation is used to compute the observable parameters.

$\bullet$ Uniqueness: From Henneaux's work \cite{Henneaux_1984}, the form of the scalar field Lagrangian in four-dimensional Lorentzian spacetime seems to be unique. 

 \noindent In the recent work \cite{PhysRevD.106.035020}, the non-uniqueness property of Lagrangian was re-interpreted to construct the non-standard Lagrangian of the complex scalar field. The total Lagrangian of the complex scalar field can be written in the linear combination between the standard and non-standard Lagrangians, which provide the same EOM in the appropriate energy limit.  This non-standard  Lagrangian is systematically solved from the inverse problem of the calculus of variations and this one is applied to explain various naturalness problems such as, hierarchy problem \cite{HeirarchyProblem, Susskind:1978ms} and the strong CP-problem, without introducing the symmetry extension or proposing the new elementary particles.  
 It is quite natural to extend the idea of non-standard Lagrangian construction to the scalar field and fermion field with an insightful mathematical trick since there are the remains of unsolved problems outside the complex scalar field.

In this paper, we apply the non-uniqueness idea to implement the non-standard Lagrangian of fermion and the classical homogeneous background scalar fields. In Sec.~II, we will show that, if the real scalar has a homogeneous property, such as the real scalar field in the homogeneous and isotropic universe in the dark energy \cite{PhysRevD.37.3406, PhysRevLett.80.1582, PhysRevLett.81.3067,CALDWELL200223, PhysRevD.68.023509} and inflation \cite{M0xi0,ARMENDARIZPICON1999209, DilatonModel,DilatonModel2,induce1,CiteTheActionOfHiggsInflation5,induce2, HiggsInflation}, the Lagrangian is not unique. The Lagrangian of real scalar field and fermion field can be written in the multiplicative form. Moreover, we are going to show that the Lagrangian of the classical homogeneous background scalar field and fermion field can be written in the form of the linear combination between standard and non-standard ones while the classical EOM is still intact. In Sec-III, some applications of the non-uniqueness property of Lagrangian in the quantum field theory and cosmology are also discussed. In Sec-IV, this section is the conclusion.

\section{Non-standard Lagrangian of the classical background scalar field from the non-uniqueness principle}

\subsection{Non-uniqueness of homogeneous real scalar field Lagrangian}
We will briefly explain the idea of the non-uniqueness principle.  In general, the relativistic motion of the real scalar field ($\phi$) is traditionally described through Klein-Gordon equation
\begin{align}\label{KG1}
\partial_\mu\partial^\mu\phi+\frac{\partial V}{\partial \phi}=0,
\end{align}
where $V$ is a scalar potential of $\phi$. In four-dimensional spacetime, the Lorentz invariant Lagrangian of this field is unique \cite{Henneaux_1984} and can be written only in the standard form as 
\begin{align}\label{Lstandard}
    \mathcal{L}_\text{Standard}=\frac{1}{2}\partial_\mu\phi\partial^\mu\phi-V.
\end{align}
In other words, there is only a single form of the scalar field Lagrangian that can provide Eq.~\eqref{KG1}. However, according to an isotropic property of the universe, the classical background value of real scalar field $\phi_b$ is spatially homogeneous. Then, the scalar field can be separated into 
\begin{align}\label{sepfield}
\phi(t,\boldsymbol{x})=\phi_b(t)+\varphi(t,\boldsymbol{x}),
\end{align}
where $\varphi(t,x)$ is the fluctuation of a quantized field, while, in cosmological fashion, $\varphi(t,x)$  is frequently called a perturbed field.  The Lagrangian \eqref{Lstandard} is rewritten as
\begin{align}
    \mathcal{L}_\text{Standard}=\mathcal{L}_\text{Classical}+\mathcal{L}_\text{Quantum}
\end{align}
where $\mathcal{L}_\text{Quantum}=\mathcal{O}(\varphi)$, and
\begin{align}\label{Lclassical} \mathcal{L}_\text{Classical}\simeq\frac{\dot{\phi_b}^2}{2}-V(\phi_b).
\end{align}
We can see that the classical Lagrangian of $\phi_b$ only depends on time while the spatial fluctuations are encoded in the Lagrangian of  $\varphi$.

By analogy $\phi_b(t)$ as $x(t)$ in the Newtonian mechanics,  it is obvious that this classical Lagrangian of the real scalar field is not unique as the Lagrangian $\eqref{LMbig1}-\eqref{LMbig2}$. Therefore, there are many non-standard Lagrangians in terms of $\phi_b$ and $\dot{\phi}_b$ providing the Klein-Gordon equation \eqref{KG1}.  

\subsection{Non-standard Lagrangian of homogeneous real scalar field}
In this contribution, we construct the non-standard Lagrangian of the scalar field by employing the inverse problem of the calculus of variations \cite{douglas1941solution,hojman1981inverse,sarlet1982helmholtz,surawuttinack2016multiplicative}. We are interested in the multiplicative form of the non-standard Lagrangian given by
\begin{align}\label{L1}
    \mathcal{L}_\text{Non-standard}=F(X)f(\phi),
\end{align}
where $\phi=\phi(t,x)$ is the real scalar field, $X=\partial_{\mu}\phi\partial^\mu \phi/2$. The functions $F$ and $f$ in Eq.~\eqref{L1} are unknown and these functions can be determined. By substituting Eq.~\eqref{sepfield} into Eq.~\eqref{L1}, the multiplicative Lagrangian can be reorganized into the classical background part and the quantum fluctuation part as 
\begin{align}\label{L01}
 \mathcal{L}_\text{Non-standard}=F\left(X_b\right)f(\phi_b)+\mathcal{O}(\varphi,\partial_\mu\varphi\partial^\mu\varphi,\dot{\varphi}).
\end{align}
where $X_b=\dot{\phi}_b^2/2$.
By ignoring the small fluctuation part $\mathcal{O}(\varphi,\partial_\mu\varphi\partial^\mu\varphi,\dot{\varphi})$ and employing the non-uniqueness, the EOM of the classical background part of Lagrangian \eqref{Lclassical} and \eqref{L01} is required to be
\begin{align}\label{KG}
    \ddot{\phi}_b+\frac{\partial V}{\partial \phi_b}=0.
\end{align}
 Substituting Eq.~\eqref{L01} into the Euler-Lagrange equation,
\begin{align}
    \frac{\partial\mathcal{L}}{\partial\phi_b}-\partial_0\frac{\partial\mathcal{L}}{\partial\dot{\phi}_b}=0,
\end{align}
we have
\begin{align}\label{EOM1}
    0=\left(F-2X_b\frac{\partial F}{\partial X_b}\right)\frac{\partial f}{\partial\phi_b}-\ddot{\phi}_b f\left(\frac{\partial F}{\partial X_b}+2X_b \frac{\partial^2 F}{\partial X_b^2}\right).
\end{align}
Then, replacing $\ddot{\phi}_b$ in the second term with the help of the Klein-Gordon equation \eqref{KG} and applying the separation variable method, Eq.~\eqref{EOM1} can be reorganized into
\begin{align}
    &\frac{d f}{d\phi_b}=-\frac{1}{\epsilon\Lambda^4}f\frac{\partial V}{\partial\phi_b},\label{SVM1}
    \\
    &2X_b\frac{d^2 F}{d X_b^2}+\frac{d F}{d X_b}=\frac{1}{\epsilon\Lambda^4}\left(F-2X_b\frac{d F}{d X_b}\right),\label{SVM2}
\end{align}
where $\Lambda^4$ is an arbitrary positive constant with mass dimension $[\Lambda]=1$ and $\epsilon=\pm 1$. We note that, mathematically, $\epsilon$ can be $\pm i$, but these complex Lagrangians are not in the scope of our study. By solving Eq.~\eqref{SVM1} and Eq.~\eqref{SVM2} together with Eq.~\eqref{L01}, the full multiplicative Lagrangian in 4-dimensional Lorentzian spacetime in Eq.~\eqref{L1} is given by
\begin{align}\label{L1sol2}
   & \mathcal{L}_\text{Non-standard}=\nonumber
    \\
    &\left( \Lambda ^2 \sqrt{\epsilon\pi X}  e^{\frac{X}{\epsilon\Lambda ^4  }} \text{erf}\left(\sqrt{\frac{X}{\epsilon\Lambda^4}}\right)+\epsilon\Lambda ^4  \right) e^{-\frac{V+X}{\epsilon\Lambda ^4  }}-\epsilon\Lambda^4.
\end{align}
The EOM of the classical background field of Lagrangian \eqref{L1sol2}  is indeed the Klein-Gordon equation, with an exponential factor, as
\begin{align}\label{EOMLm}
    0=e^{-\frac{\dot{\phi_b^2}+2V}{2\epsilon\Lambda^4}}\left(\ddot{\phi_b}+\frac{\partial V}{\partial\phi_b}\right).
\end{align}
We can conclude that the Lagrangian \eqref{L1sol2} is an alternative form to explain the relativistic motion of the homogeneous real scalar field. With this non-uniqueness property, one can choose the standard Lagrangian, non-standard Lagrangian, or both types to obtain the Klein-Gordon equation. As a consequence, the Lagrangian of classical homogeneous real scalar field can be written in the form
\begin{align}\label{Lsumassumption0}
    \mathcal{L}_{\phi}=\alpha_0 \mathcal{L}_\text{Standard}+ \alpha\mathcal{L}_{\text{Non-standard},\Lambda},
\end{align}
where $\alpha_0$ and $\alpha$ are arbitrary constant. 


\section{Non-standard Lagrangian of fermion field}

\subsection{Lagrangian}
In this section, we construct the multiplicative Lagrangian of fermion field $\psi$ coupling with a spacetime dependent current $J(x)$
\begin{align}\label{LMdirac}
\mathcal{L}=G(i\overline{\psi}\gamma^\mu\partial_\mu\psi)g(\overline{\psi}J\psi),
\end{align}
which leads to the Dirac equation as
\begin{align} \label{DiracEq}
i \gamma^\mu\partial_\mu\psi-J\psi=0,
\end{align}
where $J$ can be parametrized in the form 
\begin{align}
J(x)=a(x)+b_\mu(x)\gamma^\mu+c_\mu(x)\gamma^\mu\gamma^5.
\end{align}
Substituting Eq.~\eqref{LMdirac} into Euler-Lagrange equation
\begin{align}
    0=\frac{\partial \mathcal{L}}{\partial\overline{\psi}}-\partial_\mu\frac{\partial\mathcal{L}}{\partial \partial_\mu\overline{\psi}},
\end{align}
we have
\begin{align}\label{DiracEq1}
    0=i\gamma^\mu\partial_\mu\psi g\frac{\partial G}{\partial Y}+J\psi G\frac{\partial g}{\partial y},
\end{align}
where $Y=i\overline{\psi}\gamma^\mu\partial_\mu\psi$ and $y=\overline{\psi}J\psi$.
Then, by replacing the derivative term in Eq.~\eqref{DiracEq1} with Eq.~\eqref{DiracEq} and applying the separation variable method, this equation can be reorganized into
\begin{align}
\frac{dg}{dy}=&-\frac{1}{\epsilon\Lambda^4}g,\label{Gg}
\\
\frac{dG}{dY}=&+\frac{1}{\epsilon\Lambda^4}G,\label{Gg2}
\end{align}
where the property of $\epsilon$ and $\Lambda$ was discussed in previous section, see Eq.~\eqref{SVM1}-\eqref{SVM2}.
By solving Eq.~\eqref{Gg}-\eqref{Gg2}, the multiplicative Lagrangian for fermion field that provides the Dirac equation \eqref{DiracEq} is given by
\begin{align} \label{Lmfermion0}\mathcal{L}=\epsilon\Lambda^4e^{\frac{i\overline{\psi}\gamma^\mu\partial_\mu\psi-\overline{\psi}J\psi}{\epsilon\Lambda^4}}.
\end{align}
For the same reason as the case of the real scalar field, the Lagrangian of the fermion field can be rewritten in the form of
\begin{align}\label{Lmfermion}
    \mathcal{L}_f=\beta_0\left(i\overline{\psi}\gamma^\mu\partial_\mu\psi-\overline{\psi}J\psi\right)+\beta\epsilon\Lambda^4e^{\frac{i\overline{\psi}\gamma^\mu\partial_\mu\psi-\overline{\psi}J\psi}{\epsilon\Lambda^4}},
\end{align}
where $\beta_0$ and $\beta$ are arbitrary constant. 

We note that the Lagrangian of the fermion field processes the non-uniqueness property in the different way from the Lagrangian of the real scalar field. In the case of the real scalar field, one requires the classical background to be homogeneous, see Eq.~\eqref{sepfield}. Then, there exists an alternative form of the Lagrangian. On the other hand, in the case of the fermion field, one can solve the alternative Lagrangian without requiring a homogeneous background. Therefore, the non-uniqueness is applicable to the four-dimensional fluctuation of the spinor field, including the classical interpretation and quantization itself.

\section{Discussions}
We discuss  the contribution of our proposed non-standard Lagrangian to cosmological and particle physics.

\subsection{Ghost condensate}
\begin{figure}[H]
    \centering
    \includegraphics[width=0.35\textwidth]{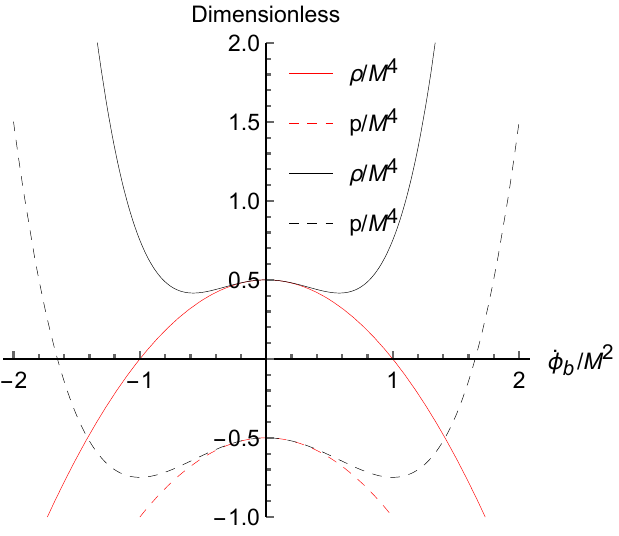}
    \caption{The sketch of the energy and pressure of phantom field where $V=0.5M^4$, and the red line and black line are of the standard phantom and the ghost condensate model, respectively}
    \label{fig:energydensityofphantom}
\end{figure}

We discuss the contribution from the non-uniqueness to be background idea of the ghost condensate model. In brief, the scalar field with negative sign kinetic energy,
\begin{align}\label{phantom}
    \mathcal{L}_\text{phantom}=-X-V,
\end{align}
is the so-called phantom or ghost field in cosmology and particle physics, respectively. This field model is frequently used to explain the dark energy with the equation of state $w<-1$ \cite{PhysRevD.68.023509, PhysRevD.68.023522, PhysRevLett.91.071301, NOJIRI2003147}. However, this model can cause the stability problem to the quantum field theory due to the unbounded energy density and pressure of the background field $\rho=-\dot{\phi}_b^2/2+V$, and $p=-\dot{\phi}_b^2/2-V$, respectively, see the red and red dashed line in Fig.~\eqref{fig:energydensityofphantom}. To get rid of the unbounded energy density, the ghost condensate is an alternative solution \cite{Arkani-Hamed:2003pdi, Krotov:2004if, Koehn:2012te} and the phantom field is considered as the effective theory. The negative kinetic energy is assumed to be the leading order approximation from the UV theory, so the negative unbounded kinetic energy can be dominated by the positive contribution at the energy scale $M$ as
\begin{align}\label{phantom2}
    \mathcal{L}_\text{phantom}=-X+\frac{X^2}{M^4}-V,
\end{align}
where $M$ is the cutoff scale of this effective theory. The energy density and pressure of the homogeneous phantom background are bounded from below with a Mexican hat-shape
\begin{align}
    \rho=-\frac{\dot{\phi}_b^2}{2}+\frac{3\dot{\phi}_b^4}{4M^4}+V,
    \\
    p=-\frac{\dot{\phi}_b^2}{2}+\frac{\dot{\phi}_b^4}{4M^4}-V,
\end{align}
see the black and black dashed lines in Fig~\ref{fig:energydensityofphantom}.
By expanding  the background phantom field around the saddle point of the Lagrangian $\phi=\langle\dot{\phi}_b\rangle t  +\varphi$, which 
$\langle\dot{\phi}_b\rangle=M^2$ is evaluated by $\partial p/\partial\dot{\phi}_b=0$, the exciting field or the ghost particle $\varphi$ is condensed into the  particle with positive kinetic energy. Therefore, the vacuum state is protected from the infinite negative energy state. In general, this framework can be obtained from several physics such as the quintom scenario \cite{Feng:2004ad, Cai:2008qb, Cai:2009zp} and the dilatonic ghost condensate \cite{Piazza:2004df} motivated from the string theory. 

\begin{figure}[H]
    \centering
  \includegraphics[width=0.35\textwidth]{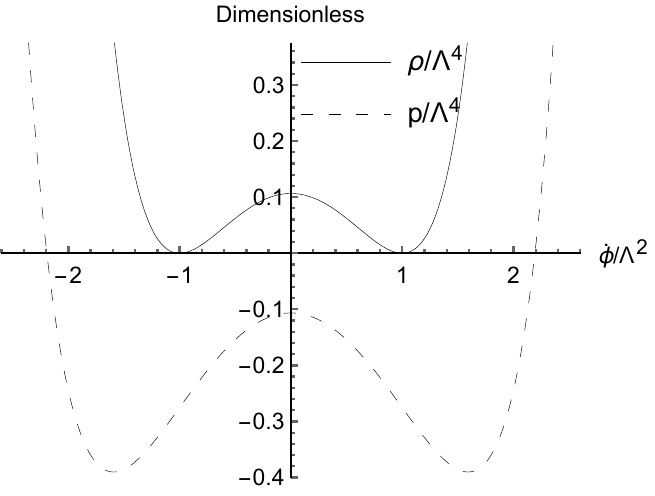}
    \caption{The shape of energy density \eqref{energydensity0} and the pressure  \eqref{pressure0} with $\epsilon=+1$ $\alpha=-1$, and $V=0.5\Lambda^4$}
    \label{fig:energydensity}
\end{figure}

In our contribution, according to an isotropic of the universe, the phantom field is obviously required to be homogeneous so the non-standard term from non-uniqueness of the Lagrangian is available with the invariant of EOM. Therefore, the Lagrangian of the phantom field can be written as
\begin{align}\label{Lphantom0}
   & \mathcal{L}_\phi=-X-V\nonumber
    \\
    &-\alpha  \Lambda ^2 \sqrt{ \pi\epsilon X } e^{-\frac{V}{\epsilon\Lambda ^4  }} \text{erfi}\left(\sqrt{\frac{X}{\epsilon \Lambda^4}}\right)+\alpha\epsilon   \Lambda ^4  e^{\frac{X-V}{\epsilon \Lambda ^4 }}-\alpha \Lambda^4,
\end{align}
where we have set $\alpha_0=1$, $\text{erfi}(x)$ is an imaginary error function defined as $\text{erfi}(x)=\text{erf}(i x)/i$. Both standard and non-standard Lagrangian give the classical EOM
\begin{align}\label{phantomEOM2}
    \left(\ddot{\phi}_b-\frac{\partial V}{\partial\phi_b}\right)=0.
\end{align}
Then, the energy density and pressure of Eq.~\eqref{Lphantom0} can be derived from
\begin{align}
\rho_\phi=&\frac{\partial\mathcal{L}_\phi}{\partial\dot{\phi}_b}\dot{\phi}_b-\mathcal{L}_\phi,\quad\text{and}\quad p_\phi=\mathcal{L}_\phi,
\end{align}
resulting in
\begin{align}
    \rho_\phi=&-\frac{\dot{\phi }_b^2}{2}+V-\epsilon\alpha\Lambda^4 e^{\frac{\dot{\phi }_b^2-2 V}{2\epsilon \Lambda ^4}}+\alpha \Lambda^4, \label{energydensity0}
    \\
    p_\phi=&-\alpha  \Lambda ^4-\sqrt{\frac{\pi\epsilon }{2}} \alpha  \Lambda ^2  \dot{\phi }_b e^{-\frac{V}{\epsilon\Lambda ^4  }}
   \text{erfi}\left(\frac{\dot{\phi }_b}{\sqrt{2\epsilon} \Lambda ^2 }\right)\nonumber
   \\
   &+\epsilon\alpha  \Lambda ^4   e^{\frac{\dot{\phi }_b^2-2 V}{2 \epsilon\Lambda ^4
    }}-V-\frac{\dot{\phi }_b^2}{2}.\label{pressure0}
\end{align}
The behavior of the energy density and pressure depends on  $\epsilon$ and $\alpha$. We find that there is one interesting choice. If $\epsilon=+1$ and $\alpha=-1$,  the exponential kinetic energy in the third term can provide the positive contribution to the energy density and pressure of the phantom field in the limit $\dot{\phi}\gg\Lambda^2$. The energy density and the pressure are now bounded from below, sketched in Fig.~\ref{fig:energydensity}. 
Therefore, the ground state of the ghost can be protected, and the ghost condensate can naturally exist with the multiplicative Lagrangian.

\subsection{Unnatural smallness of cosmological constant}
The unnatural small scale of the cosmological constant \cite{aghanim2020planck},
\begin{align}
    \Lambda_{cc}\simeq 10^{-47}\text{GeV}^4,
\end{align}
 comparing to the particle physics scale and the Planck scale is one of the questionable problems \cite{RevModPhys.61.1,carroll2001cosmological, LOMBRISER2019134804,Shapiro:2000dz, Shapiro:1999zt,EduardV.Gorbar_2004, Sola:2013gha,Sola:2014tta} from the naturalness principle \cite{Williams:2015gxa,rosaler2019naturalness,vanKolck:2020plz, Susskind:1978ms}.  The dimensionless ratio of parameters in the model is expected around order unity $\mathcal{O}(1)\sim 1-10^{-3}$ or $\mathcal{O}(1)\sim 1-10^{-3}$ \cite{Dirac:1938mt}. If the violation occurs, it leads to the fine-tuning problem and the existence of new physics. The ratio between 
electroweak vacuum energy $V_\text{EW}\simeq 10^8$ GeV$^4$ and the cosmological constant is around fifty order of magnitude
\begin{align}
    \frac{V_\text{EW}}{\Lambda_{cc}}\sim 10^{55}.
\end{align}
where $V_\text{EW}$ is obtained by the electroweak phase transition from the Higgs potential, 
\begin{align}
    V=&-\frac{m_h^2}{2}H^\dagger H+\lambda_h (H^\dagger H)^2,
    \\
    =&\frac{m_h^2v_h^2}{4}+\frac{m_h^2}{2}h^2+O(h^3),
    \\
    =&V_\text{EW}+\frac{m_h^2}{2}h^2++O(h^3),
\end{align}
which $m_h=125$GeV is Higgs mass, $v_h=246$GeV is Higgs vacuum expectation value (VEV) \cite{Erler:2019hds}, and $h$ is quantum fluctuation of Higgs field. The artificial constant term $V_0$ is required to accurately fine-tune $\Lambda_{cc}=V_\text{EW}+V_0$ around 55 orders of decimal. This unnatural situation is displeasure to some physicists in the community. Setting $V_0=-V_\text{EW}$ is more natural way, meanwhile, a new physical reason behind this fine-tuning comes from the new physics $\Lambda_{cc}=V_\text{EW}+V_0+V_{new}$. For instance, if there is the new particle spontaneously breaking down to the minimum at $V_{new}\sim m^2v^2\sim m^4/\lambda\sim 10^{-47}(\text{GeV})^4$, the induced scale of the cosmological constant is possible. However, the unnatural situation still appears, because the coupling constant $\lambda$ is extremely large or a mass parameter $m$  has to be incredibly smaller than the SM particle.  The status of naturalness is still distress that why this particle has a very large dimensionless coupling compared with the gauge coupling and Yukawa coupling of SM particle.

Here, in our contribution, we consider the real scalar field in the Ginzburg Landau potential
\begin{align}
    \mathcal{L}=X+\frac{m^2_\phi}{2}\phi^2-\frac{\lambda_\phi}{4}\phi^4
\end{align}
which $\phi$ can be separated into the classical homogeneous background and the fluctuation field as $\phi(x)=\phi_b(t)+\varphi(t,\boldsymbol{x})$. The classical background $\phi_b$ satisfy the EOM
\begin{align}\label{phi4EOM}
    \ddot{\phi}_b-m_\phi^2\phi_b+\lambda_\phi \phi_b^3=0.
\end{align}
The vacuum solution of field can be obtained by setting $\dot{\phi}_b=0$. Then, we have
\begin{align}\label{phi4solution}
    \langle\phi_b\rangle=v_\phi=\frac{m_\phi}{\sqrt{\lambda_\phi}}.
\end{align}
This solution relates with the ground state from the classical energy density
\begin{align}
    \rho=\frac{\dot{\phi}_b^2}{2}-\frac{m^2_\phi}{2}\phi_b^2+\frac{\lambda_\phi}{4}\phi_b^4.
\end{align}
 At late time $t\to\infty$, the real scalar field evolves to the lowest energy state at 
\begin{align}\label{shiftfield4}
    \phi=v_\phi+\varphi
\end{align}
with the remains of small fluctuation $\varphi$.

\begin{figure}[H]
      \centering
      \includegraphics[width=0.45\textwidth]{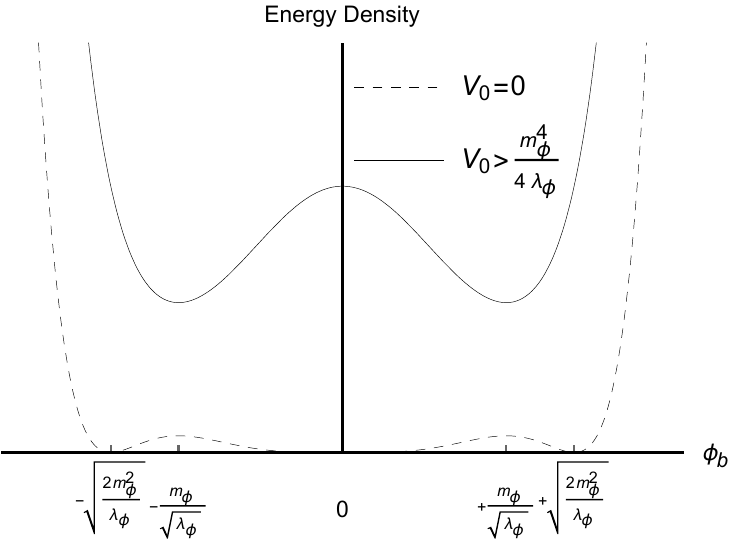}
      \caption{The energy density of Eq.~\eqref{phi4MLagrangian1} in the small kinetic energy of the background field $\dot{\phi_b}\ll\Lambda^2$} 
      \label{fig:unstablepotential}
  \end{figure}

Here, according to the homogeneous background in the classical part, the non-uniqueness is applicable to write the multiplicative Lagrangian as
\begin{align}\label{phi4MLagrangian}
  &\mathcal{L}=X+\frac{m^2_\phi}{2}\phi^2-\frac{\lambda_\phi}{4}\phi^4-\Lambda ^4\nonumber
  \\
    &- \Lambda ^2 \sqrt{\pi X} \text{erfi}\left(\frac{\sqrt{X}}{\Lambda ^2}\right) e^{\frac{\frac{\lambda_\phi  \phi ^4}{4}-\frac{m_\phi ^2 \phi
   ^2}{2}}{\Lambda ^4}}+\Lambda ^4 e^{\frac{\frac{\lambda_\phi  \phi ^4}{4}-\frac{m_\phi^2 \phi
   ^2}{2}+X}{\Lambda ^4}},
\end{align}
where, we have chosen $\alpha_0=1$, $\alpha=-1$, and $\epsilon=1$.
 The EOM of Eq.~\eqref{phi4MLagrangian} still satisfies Eq.~\eqref{phi4EOM}. This means that the vacuum solution of $\phi_b$ is still obtained in the same form as Eq.~\eqref{phi4solution}. However, there is a problem in the following. By considering energy density of Eq.~\eqref{phi4MLagrangian} with $\partial_\mu\phi=0$, there are four-extrema points with
 \begin{align}
     \phi=0,\pm\frac{m_\phi}{\sqrt{\lambda_\phi}},\pm \frac{\sqrt{2}m_\phi}{\sqrt{\lambda_\phi}}.
 \end{align}
  We find that $v_\phi=m_\phi/\sqrt{\lambda_\phi}$ is no longer the ground state of theory, see the black dashed line in Fig.~\ref{fig:unstablepotential}. This means that vacuum solution at $\phi_b=m_\phi/\sqrt{\lambda_\phi}$ does not lead to the ground state of theory. The expansion of field around $\phi_b=m_\phi/\sqrt{\lambda_\phi}$ is not reasonable.

We fix this problem by adding the constant term $V_0$ into the potential $V$, which does not change the classical EOM. The Lagrangian \eqref{phi4MLagrangian} is rewritten as
\begin{align}\label{phi4MLagrangian1}
  &\mathcal{L}=X-V_0+\frac{1}{2}m^2_\phi\phi^2-\frac{\lambda}{4}\phi^4-\Lambda ^4\nonumber
  \\
    &- \Lambda ^2 \sqrt{\pi X} \text{erfi}\left(\frac{\sqrt{X}}{\Lambda ^2}\right) e^{\frac{\frac{\lambda  \phi ^4}{4}-\frac{\mu ^2 \phi
   ^2}{2}+V_0}{\Lambda ^4}}\nonumber
   \\
   &+\Lambda ^4 e^{\frac{\frac{\lambda  \phi ^4}{4}-\frac{\mu ^2 \phi
   ^2}{2}+V_0}{\Lambda ^4}+\frac{X}{\Lambda ^4}}.
\end{align}
We find that the extrema points of energy density of Eq.~\eqref{phi4MLagrangian1} are modified as
\begin{align}\label{ground6}
    \phi=&0,~\pm\frac{m_\phi}{\sqrt{\lambda_\phi}},~\pm\sqrt{\frac{m_\phi^2}{\lambda_\phi}-\frac{\sqrt{m_\phi^4-4V_0\lambda_\phi}}{\lambda_\phi}},\nonumber
    \\
    &~\pm\sqrt{\frac{m_\phi^2}{\lambda_\phi}+\frac{\sqrt{m_\phi^4-4V_0\lambda_\phi}}{\lambda_\phi}}.
\end{align}
We notice that if
 \begin{align}
     V_0>\frac{m_\phi^4}{4\lambda_\phi},
 \end{align}
 the square root in the third and fourth terms of Eq.~\eqref{ground6} turns to imaginary value so
 the ground state of theory is returned to be $v_\phi=m_\phi/\sqrt{\lambda_\phi}$, see the solid black line in Fig.~\ref{fig:unstablepotential}. Here, the constant in potential $V$  must be carefully chosen to protect the ground state of theory.
 Then, we can apply Eq.~\eqref{shiftfield4} into the Lagrangian \eqref{phi4MLagrangian1}. The leading  Lagrangian in terms of quadratic $\varphi$ is given by
\begin{align}\label{phi4MLagrangian2}
    &\mathcal{L}=-\Lambda ^4 \left(1-e^{-\frac{(a-1)M_{\phi }^2 v_{\phi }^2}{8
   \Lambda ^4}}\right)+\frac{1}{8} M_{\phi }^2 v_{\phi }^2\nonumber
    \\
    &+\left(1-e^{-\frac{(a-1)M_{\phi }^2 v_{\phi }^2}{8 \Lambda ^4}}\right)\left(\frac{\partial_\mu\varphi\partial^\mu\varphi}{2}-\frac{1}{2}  M_{\phi }^2 \varphi ^2\right)+\mathcal{O}(\varphi^3),
\end{align}
where $M_\varphi=\sqrt{2}m_\varphi$ is a mass of particle $\varphi$ defined from the pole of tree-level propagator and we have parametrized $V_0=a M_\phi^4/\lambda_\phi$. To preserve the naturalness, we need $a\simeq O(1)$ and the dimensionless ratios between the parameter of this scalar boson and Higgs boson are of order unity
\begin{align}\label{naturalness1}
    \frac{m_\phi}{m_h}\sim\mathcal{O}(1),~\frac{v_\phi}{v_h}\sim\mathcal{O}(1),
\end{align}
which lead to $\lambda_\phi/\lambda_h\sim \mathcal{O}(1)$, $10^2\text{GeV}<m_\phi<10^2$ TeV, and $10^2\text{GeV}<v_\phi<10^2$ TeV.  If the SM cutoff is assumed to be the Planck scale, it is appropriate to set
\begin{align}\label{Planckscale}
    \Lambda=M_p
\end{align}
because $\Lambda$ is an expansion parameter of the multiplicative Lagrangian leading to the effective interaction $\varphi^6/M_p^2$. According to Eq.~\eqref{naturalness1}-\eqref{Planckscale}, we have $M_\phi^2v_\phi^2/\Lambda^4\ll 1$ so the leading order of
the cosmological constant term from Lagrangian \eqref{phi4MLagrangian2} is suppressed by $M_p^4$ as
\begin{align}
    \Lambda ^4 \left(e^{-\frac{(a-1)M_{\phi }^2 v_{\phi }^2}{8
   \Lambda ^4}}-1\right)+\frac{1}{8}(a-1) M_{\phi }^2 v_{\phi }^2\simeq\frac{(a-1)^2M_\phi^4v_\phi^4}{128M_p^4}.
\end{align}
By ignoring the constant factor $(a-1)^2\sim \mathcal{O}(1)$, we find that if the scalar particle $\varphi$ has $v_\phi\simeq M_\phi\simeq 4$TeV, we obtain
\begin{align}
   \frac{M_\phi^4v_\phi^4}{128M_p^4}\sim 10^{-47} \text{GeV}^4.
\end{align}
Therefore, an unnatural small scale of the cosmological constant can be induced from the scalar field with a mass in TeV if the multiplicative Lagrangian obtained from the non-uniqueness property is a part of particle and field theory.  No new scalar field with an incredibly small mass and unnaturally large coupling constant is required.

\subsection{QED Lagrangian with multiplicative Lagrangian of fermion}
We discuss the modification of quantum electrodynamics (QED) due to the non-uniqueness Lagrangian of the fermion field. By considering,
\begin{align}
    a(x)=m, ~b_\mu(x)=e A_\mu(x), ~c_\mu\gamma^\mu\gamma^5=0,
\end{align}
the Lagrangian \eqref{Lmfermion} becomes
\begin{align}
&\mathcal{L}_f=\beta_0\left(i\overline{\psi}\gamma^\mu\partial_\mu\psi-m\overline{\psi}\psi-eA_\mu\overline{\psi}\gamma^\mu\psi\right)\nonumber
\\
&+\beta\epsilon\Lambda^4e^{\frac{i\overline{\psi}\gamma^\mu\partial_\mu\psi-m\overline{\psi}\psi-eA_\mu\overline{\psi}\gamma^\mu\psi}{\epsilon\Lambda^4}}.
\end{align}
This Lagrangian is invariant under local U(1) gauge transformation
\begin{align}
    \psi(x)\to e^{-i\theta(x)}\psi(x),~A_\mu(x)\to A_\mu(x)+\frac{1}{e}\partial_\mu\theta(x).
\end{align}
The modified QED Lagrangian can be constructed by adding the dynamical term of $A_\mu$ field
\begin{align}
\mathcal{L}_\text{QED}=\mathcal{L}_f-\frac{1}{4}F_{\mu\nu}F^{\mu\nu}.
\end{align}
where $F_{\mu\nu}=\partial_\mu A_\nu-\partial_\nu A_\mu$ and the condition $\alpha=1-\alpha_0$ is required to provide the canonical kinetic energy in low energy limit. In the limit 
\begin{align}
\overline{\psi}\gamma^\mu\partial_\mu\psi\ll\Lambda^4,~m\overline{\psi}\psi\ll \Lambda^4,~eA_\mu\overline{\psi}\gamma^\mu\psi\ll\Lambda^4,
\end{align}
the non-renormalized interactions up to $1/\Lambda^{4}$  can be generated
\begin{align}
O_6=&\frac{m^2}{2\Lambda^4}\left(\overline{\psi}\psi\right)^2,\label{O6}
\\
O_7=&\frac{me}{2\Lambda^4}\overline{\psi}\psi A_\nu\overline{\psi}\gamma^\nu\psi,~\frac{m}{4\Lambda^4}\overline{\psi}\psi\overline{\psi}\gamma^\mu\partial_\mu\psi,\label{O7}
\\
    O_8=&\frac{e}{2\Lambda^4}A_\nu\overline{\psi}\gamma^\nu\psi\overline{\psi}\gamma^\mu\partial_\mu\psi, ~\frac{e^2}{4\Lambda^4}\left(A_\nu\overline{\psi}\gamma^\nu\psi\right)^2,\nonumber
    \\
    &~\frac{1}{4\Lambda^4}\left(\overline{\psi}\gamma^\mu\partial_\mu\psi\right)^2,\label{O8}
\end{align}
where we ignore the factor $(1-\beta_0)/\epsilon$, which is multiplied to all theses terms.
The energy scale of $\Lambda$ can be an arbitrary value. 

$\bullet\Lambda=$Landua-pole of QED: The non-renormalized operator from non-uniqueness is very small to observe in the experiment, such as the particle collider and the g-factor experiment.

$\bullet\Lambda=M_p=2.44 \times 10^{18}$GeV: The perturbative unitary of QED scattering amplitude can break down at the scale of gravity. The effect of these non-renormalized operators from this modified QED has to be incorporated with the non-renormalized interaction of graviton. However, it is still impossible to observe in any current experiment.

$\bullet \text{TeV}<\Lambda<M_p:$ If $\Lambda$ is in this range, the effect of the higher dimension operator \eqref{O6}-\eqref{O8} from the non-uniqueness is possible to observe in the current particle collider experiment with 13 TeV collision at the large hadron collider (LHC) or the future particle collider such as the high luminosity large hadron collider (HL-LHC).

$\bullet \Lambda<\text{TeV}$: This regime of $\Lambda$ is evidently ruled out. The observed quantities such as the electron factor can modified by the effective interaction 
\begin{align}
    e\frac{m}{\Lambda^4}A_\mu\overline{\psi}\gamma^\mu\psi\overline{\psi}\psi
\end{align}
at the one-loop level with contribution $\sqrt{\alpha}m^4/\Lambda^4$, which leads to large modification on the QED prediction. Moreover, the four-fermion interactions from the operators \eqref{O6}-\eqref{O8} can address the large modification of the electron-positron scattering and the operators, including $A\psi\psi\psi\psi$, could lead to the large one-loop correction to the pair production process, which has never been reported in the large-electron positron collider (LEP) \cite{L3:1990jpg, L3:1992elb, L3:1995nbq, Burch:2005ts}.

\subsection{Chiral symmetry breaking}
We discuss the feature of multiplicative Lagrangian to induce the non-zero vacuum expectation value of the bilinear fermion operator in the explicit chiral symmetry breaking.
Let us consider the toy model of the fermion field in multiplicative form  coupled with complex scalar particle
\begin{align}
\mathcal{L}=&\beta_0\left(i\overline{\psi}\gamma^\mu\partial_\mu\psi-y\phi\overline{\psi}\psi\right)-\beta\Lambda^4e^{\frac{i\overline{\psi}\gamma^\mu\partial_\mu\psi-y\psi\overline{\psi}\psi}{\epsilon\Lambda^4}}\nonumber
\\
&+\partial_\mu\phi^*\partial^\mu\phi+m\phi^*\phi-\lambda (\phi^*\phi)^2,
\end{align}
where we have defined $\epsilon=-1$,~ $a(x)=y\phi(x)$, $\gamma_\mu a^\mu=0$, and $\gamma_\mu c^\mu=0$. Here,  the vacuum expectation value of $\phi$ is
 \begin{align}
     \langle\phi^*\phi\rangle=\frac{\mu^2}{\lambda}=v^2,
 \end{align}
so the symmetry breaking of $\phi$ is required. With ignoring the Goldstone boson and scalar field, the fermion Lagrangian is
\begin{align}
\mathcal{L}_f=&\beta_0\left(i\overline{\psi}\gamma^\mu\partial_\mu\psi-yv\overline{\psi}\psi\right)-\beta\Lambda^4e^{-\frac{i\overline{\psi}\gamma^\mu\partial_\mu\psi-yv\overline{\psi}\psi}{\Lambda^4}}.
\end{align}
A mass of fermion can be defined from the EOM as 
\begin{align}
    m_f=vy.
\end{align}
Then,  we perform the field redefinition
\begin{align}
    \psi\to\frac{\psi}{\sqrt{\beta+\beta_0}}
\end{align}
to get canonical kinetic energy of fermion in low-energy limit.

$\bullet$ In the large momentum limit $i\overline{\psi}\gamma^\mu\partial_\mu\psi\gg  m_f\overline{\psi}\psi$, and $i\overline{\psi}\gamma^\mu\partial_\mu\psi\gg\Lambda^4$, the exponential term vanishes so the Lagrangian can be reduced into renormalized Lagrangian containing with a pure kinetic energy
\begin{align}
    \mathcal{L}_f\simeq \frac{\beta_0}{\beta_0+\beta} i\overline{\psi}\gamma^\mu\partial_\mu\psi.
\end{align}
Therefore, the chiral symmetry is restored.

\begin{figure}[H]
     \centering
 \includegraphics[width=0.35\textwidth]{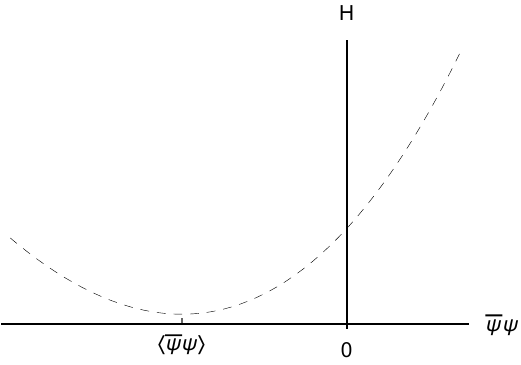}
     \caption{The sketch of Eq.~\eqref{energyf} with $\beta_0<0$}
     \label{fig:fermion}
 \end{figure}

$\bullet$ On the other hand, in the small momentum limit $m_f\overline{\psi}\psi\gg i\overline{\psi}\gamma^\mu\partial_\mu\psi$, the energy density of this fermion approximately contains only mass term 
\begin{align}\label{energyf}
    \mathcal{H}\simeq \frac{\beta_0}{\beta_0+\beta}m_f\overline{\psi}\psi+\beta\Lambda^4 e^{\frac{m_f\overline{\psi}\psi}{(\beta_0+\beta)\Lambda^4}},
\end{align}
which explicitly breaks the chiral symmetry. To make the energy density bounded from below and the kinetic energy of fermion in high energy limit be positive, the conditions
\begin{align}
    \Lambda^4>0,~m_f>0,~\beta_0<0,~\beta<-\beta_0
\end{align}
are required. We find that the ground state of the bilinear fermion operator does not appear at $\langle \overline{\psi}\psi\rangle\neq 0$ but, see Fig.~\ref{fig:fermion},
\begin{align}
    \langle \overline{\psi}\psi\rangle=\left(\beta_0+\beta\right)\frac{\Lambda^4}{m_f}\log\left(-\frac{\beta_0}{\beta}\right).
\end{align}
 With this non-zero value of the bilinear fermion operator,  the chiral condensate can happen. Therefore, if the fermion field Lagrangian is written in the standard and multiplicative form, the chiral condensate can be triggered by spontaneous symmetry breaking of scalar particle.

In our opinion, it is interesting to incorporate and extend this fermion Lagrangian into the quark model with $SU(3)$ symmetry in quantum chromodynamics. If $\Lambda^4/m_f$ is set at the scale $\Lambda_\text{QCD}\sim 200$ MeV, we can have the quark field that condenses into the pion field with non-zero VEV in the low energy and behaves like renormalized QCD with the restored chiral symmetry in the high momentum mode.

\subsection{Infinite form of Lagrangian}
In the last part of the discussion, we show that not only the multiplicative form of the Lagrangian can contribute to the Lagrangian of the homogeneous real scalar field and the fermion field but there can be  infinite polynomial forms of the non-standard Lagrangian. In the case of the homogeneous scalar field, by expanding the Lagrangian \eqref{L1sol2} in the large value of $\Lambda$, we have
\begin{align}
    \mathcal{L}_{\text{non-standard},\phi}=\mathcal{L}_\phi^{(0)}+\mathcal{L}_\phi^{(4)}+\mathcal{L}^{(8)}_\phi+...,
\end{align}
where
\begin{align}
   & \mathcal{L}_\phi^{(0)}=X-V,\label{polyphi1}
    \\
    & \mathcal{L}_\phi^{(4)}=\frac{3 V^2-6 V X-X^2}{6 \Lambda ^4 \epsilon },\label{polyphi2}
     \\
     & \mathcal{L}_\phi^{(8)}=\frac{-5 V^3+15 V^2 X+5 V X^2+X^3}{30 \Lambda ^8 \epsilon ^2}.\label{polyphi3}
\end{align}
In the case of the fermion field, by expanding the Lagrangian \eqref{Lmfermion0} in the large value of $\Lambda$, we have
\begin{align}
    \mathcal{L}_{\text{non-standard},\psi}=\mathcal{L}_\psi^{(0)}+\mathcal{L}_\psi^{(4)}+\mathcal{L}_\psi^{(8)}+...,
\end{align}
where
\begin{align}
&\mathcal{L}_\psi^{(0)}=i\overline{\psi}\gamma^\mu\partial_\mu\psi-\overline{\psi}J\psi,\label{polypsi1}
    \\
     &\mathcal{L}_\psi^{(4)}=\frac{\left(i\overline{\psi}\gamma^\mu\partial_\mu\psi-\overline{\psi}J\psi\right)^2}{2\epsilon\Lambda^4},\label{polypsi2}
     \\
      &\mathcal{L}_\psi^{(8)}=\frac{\left(i\overline{\psi}\gamma^\mu\partial_\mu\psi-\overline{\psi}J\psi\right)^3}{6\epsilon^2\Lambda^8}.\label{polypsi3}
\end{align}
The EOM of \eqref{polyphi1}-\eqref{polyphi3} and \eqref{polypsi1}-\eqref{polypsi3} can lead to the homogeneous Klein-Gordon equation and the Dirac equation, respectively, as
\begin{align}
   & \frac{1}{n!}\left(\frac{1}{\epsilon\Lambda^4}\left(\frac{\dot{\phi}_b^2}{2}+V\right)\right)^n\left(\ddot{\phi_b}+\frac{\partial V}{\partial\phi_b}\right)=0,
    \\
    & \frac{1}{n!}\left(\frac{1}{\epsilon\Lambda^4}\left(i\overline{\psi}\gamma^\mu\partial_\mu\psi-\overline{\psi}J\psi\right)\right)^n\left(i\gamma^\mu\partial_\mu\psi-J\psi\right)=0,
\end{align}
where $n=0,1,2,...$ .
According to the non-uniqueness property of Lagrangian, it is possible to write the Lagrangian of $x=(\phi,\psi)$ in the form of linear combination between standard, non-standard, and polynomial Lagrangian with the different value of $\Lambda$ as
\begin{align}
    &\mathcal{L}_x=\alpha_0\mathcal{L}_\text{STD} +\alpha\mathcal{L}_{\text{NSTD},\Lambda}+\sum_i \alpha_i\mathcal{L}_{\text{Polynomial},\Lambda_i}
\end{align}
where $\alpha_i$ is an arbitrary constant. This way of constructing the Lagrangian may lead to a new insight of unsolved problems in physics.

\section{Conclusion}
The non-uniqueness principle can lead to the modification  of the Lagrangian  in various approaches.

$\bullet$ The complex scalar field\cite{PhysRevD.106.035020}: Many non-standard Lagrangians yield the Klein-Gordon equation in the appropriate limit. Thus, within this limit, the Lagrangian is not unique.

$\bullet$ The real scalar field:  By employing the homogeneous property of the classical background scalar field, the classical background Lagrangian is not unique.

$\bullet$ The fermion field: The fermion Lagrangian is explicitly non-uniqueness in four-dimensional spacetime.

Therefore, the linear-combination Lagrangian can be realized with the invariant of the EOM.  Various phenomena in particle physics and cosmology, for example, ghost condensate, cosmological constant, and chiral condensate can be explored.

\begin{acknowledgements}
We acknowledge the support from the Petchra Prajomklao Ph.D. Research Scholarship from King Mongkut’s University of Technology Thonburi (KMUTT). 
\end{acknowledgements}

\bibliographystyle{apsrev4-1}
\bibliography{mybib.bib}
\end{document}